\documentclass[twocolumn,showpacs,preprintnumbers,amsmath,amssymb]{revtex4}

\usepackage{graphicx}
\usepackage{dcolumn}
\usepackage{bm}
\newcommand{\etaup}{\raisebox{0.2em}{$\eta$}}
\newcommand{\chiup}{\raisebox{0.2em}{$\chi$}}

\begin{document}

\preprint{}

\title{The Relativistically Spinning Charged Sphere}

\author{D. Lynden-Bell}
\affiliation{%
The Institute of Astronomy, The Observatories,\\
Madingley Road, Cambridge, CB3 0HA, UK\\
\& Clare College, Cambridge
}%

\date{\today}

\begin{abstract}
When the equatorial spin velocity, $v$, of a charged conducting
sphere approaches $c$, the Lorentz force causes a remarkable
rearrangement of the total charge $q$.
  
Charge of that sign is confined to a narrow equatorial belt at
latitudes $b \leqslant \sqrt{3} \left(1 - v^2\!/c^2
\right)^{\frac{1}{2}}$ while charge of the opposite sign occupies
most of the sphere's surface.  The change in field structure is shown
to be a growing contribution of the `magic' electromagnetic field of
the charged Kerr-Newman black hole with Newton's G set to zero. The
total charge within the narrow equatorial belt grows as $\left(
1-v^2\!/c^2 \right)^{-\frac{1}{4}}$ and tends to infinity as $v$
approaches $c$.  The electromagnetic field, Poynting vector, field
angular momentum and field energy are calculated for these
configurations.

Gyromagnetic ratio, g-factor and electromagnetic mass are illustrated
in terms of a 19th Century electron model.  Classical models with no
spin had the small classical electron radius $e^2/mc^2\sim$ a
hundredth of the Compton wavelength, but models with spin take that
larger size but are so relativistically concentrated to the equator
that most of their mass is electromagnetic.

The method of images at inverse points of the sphere is shown to
extend to charges at points with imaginary co-ordinates. 
\end{abstract}

\pacs{Valid PACS appear here}
\maketitle

\section{\label{sec:introduction}Introduction}

When a charged conductor rotates steadily, any small resistivity in the
conductor will cause dissipation, unless the charge rotates with the
conductor.  Thus, in the steady state the surface current is due to
the rotation of the surface charge.  For a sphere rotating with
equatorial speed $v$ much less than $c$, the charge is almost
uniformly distributed over the surface and the resulting magnetic
field is almost uniform within the sphere and externally dipolar.  If
the charge were frozen to be uniform over the surface and the sphere
were an insulator then those would be the forms of the fields, however
fast the sphere rotated.  (Appendix 1).

By contrast, when the sphere has a conducting surface the Lorentz force
pushes the charge along the surface towards the equator, but the effect
is only pronounced at relativistic speeds.  Beyond $v= 0.93 c$ the
effect is so strong that the charge density at the poles is of the
opposite sign to the imposed total charge $q$.  As $v$ gets still
nearer to $c$ more and more of the sphere is covered by such reversed
charge, while the charge of the same sign as $q$ is confined to an ever
narrowing belt around the equator.  However this belt now contains a
charge considerable greater than $q$ since the excess is needed to
cancel the reversed charge elsewhere.  Indeed we show that as $v$
approaches $c$ the total charge in the belt becomes $\frac{q}{2
  \Delta} \left(1- v^2\!/c^2 \right)^{-\frac{1}{4}}$ which tends to
  infinity.  The boundaries of the equatorial belt are at latitudes
  $|b|\simeq \sqrt{3}~ \sqrt{1-v^2\!/c^2}$ so the width of the belt
  tends to zero.

Defining $u$ by $(1-u)/(1+u) = \sqrt{1-v^2\!/c^2}$ we find that as $v$
increases, the external field is changed by the addition of the
electromagnetic field of a charge at the imaginary point $(0,0,ia
\sqrt{u})$.  This is the Magic electromagnetic field discussed earlier
(Newman, 1973, 2002; Lynden-Bell, 2003) which may also be found as the
field left by a charged Kerr-Newman black hole (Newman {\it et al.},
1965) when $G$ is set equal to zero.  In our context it is being
generated quite naturally as $v \rightarrow c$.  The corresponding
internal field is that due to a charge at the complex image point
under inversion in the sphere $(0,0,-ia/ \sqrt{u})$.  Wave equations
for particle motion in the Magic field have remarkable separability
(Carter, 1968b; Chandrasekhar, 1976; Page, 1976; Lynden-Bell, 2000).
All the Kerr-Newman metrics have the same gyromagnetic ratio as the
Dirac electron (Carter, 1968a) and so do all the conformastationary
metrics and others remarked upon by Pfister and King (2002).

There are several points of principle that arise in this problem which
we raise but do not discuss in detail.

\begin{enumerate}
\item 
If a real physical sphere with a conducting surface is rotated
relativistically, what shape would it become?  One might suppose that
it ought to be a sphere in the rotating axes, but the metric in such
axes is anisotropic, so it is not clear what shape should be called
spherical in such axes.  We sidestep all such discussion by taking the
surface to be spherical in fixed axes however fast the body is
rotated.  By this we elimiate all discussion of Fitzgerald contraction,
the physical constitution of the sphere and its distortion under
centrifugal and Lorentz forces.  While this may reduce the physical
reality of the problem, it gives a well defined uncomplicated problem.

\item
For a slowly rotating conductor the charge resides on the surface so
it does not matter whether the whole body of the sphere is a conductor
or only a surface layer.  However, once the magnetic part of the
Lorentz force is significant, the condition that the Lorentz force
density must vanish inside the sphere leads to a requirement for
internal charges and the currents due to their rotation.  Thus in the
relativistic r\'{e}gime the conducting solid sphere gives a problem
that differs from the conducting spherical shell problem solved here.
                                                                     
\item
We do not assume that our conductor has zero resistivity, indeed we
rely on some resistivity to ensure that in the steady state the
surface charge rotates with the body.  In particular, we do not assume
that either the surface or the body of the sphere is a superconductor,
so questions of the expulsion of magnetic flux by the Meissner effect
do not occur in our problem.

\item
A number of very interesting problems arise if it is assumed that the
charge carriers in the conductor have mass, because at relativistic
speeds the inertia of the charge carriers will itself cause them to
concentrate toward the equator.  However the Maxwell equations
themseves already account for that part of the inertia contained in
the electromagnetic field.  When an electron is accelerated the
electromagnetic field is accelerated with it, so at least part of its
mass is due to the inertia of its field energy.  To put in the full
mass of the electron as well as the Maxwell stresses of the field
distribution would involve some double counting, not to mention the
larger effect of inertia on the heavy nuclei that form the conductor's
lattice.  We shall again sidestep these problems by assuming that the
Lorentz force component along the conducting surface is zero, so that
there is no effect of inertial forces on the charge carriers (other
than those on their electromagnetic fields as implied by Maxwell's
equations).
                                                                       
\end{enumerate}

\section{\label{sec:maxeq}The Maxwell Equations for the Problem}

We call the radius of the sphere $a$ and rotate it with angular
velocity $\Omega$ so $\Omega a = v$.  We shall find it convenient to
use $\omega = \Omega a/c = v/c$.  In the steady state, the Maxwell
equations reduce to $Curl~E=0,~{\rm{div}}E=4\pi \rho,~{\rm{div}}B=0$
and $CurlB=4 \pi j$.  On the sphere's surface there is a surface
charge $\sigma(\theta)$ and a sheet current $J_{\phi} = \sigma \Omega
ac^{-1} \sin\theta$ but away from that surface there are neither
charges nor currents so we may write ${\bf{E}}= - \nabla \Phi$ and
${\bf{B}}= - \nabla \chiup $ where $\nabla^2 \Phi =0 =
\nabla^2\chiup$. The $z \rightarrow -z$ symmetry shows
${\bf{E}}$ even and ${\bf{B}}$ odd.

The solutions to Laplace's equation obeying the boundary conditions at
$r=0$ and $\infty$ are with $\mu=\cos\theta$

\begin{equation*}
\Phi = \frac{q}{a}
\left\{
\begin{array}{llll}
\displaystyle\sum\limits_{n}\left(\frac{a}{r}\right)^{2n+1} 
  \Phi_{2n} P_{2n} \left ( \mu \right)
& \quad  & 
r \geqslant a\\
\displaystyle \sum\limits_{n}\left( \frac{r}{a} \right ) ^{2n\phantom{+1}}  \Phi_{2n} P_{2n}
  \left( \mu \right )
& \quad &
r \leqslant a~,
\end{array}
\right.
\end{equation*}
and

\begin{equation*}
\chiup = \frac{q}{a}
\left \{
\begin{array}{lll}
\displaystyle\phantom{-}\sum\limits_{n} \frac{1}{2n+2}
\left(\frac{a}{r}\right)^{2n+2} \chiup_{2n+1} P_{2n +1}\left( \mu
\right) & \quad & r > a\\
\displaystyle-\sum_{n} \frac{1}{2n+1} \left( \frac{r}{a}\right )^{2n+1}  
 \chiup_{2n+1} P_{2n+1} \left( \mu \right) 
& \quad & 
r < a~.
\end{array}
\right.
\end{equation*}

Here $q$ is the total charge on the sphere and the co-efficients have
been arranged so that $\Phi$ and $\partial \chi/\partial r$ are
both continuous across $r=a$.  This corresponds to the requirements
that the surface component of {\bf{E}} and the normal component of
{\bf{B}} must be continuous.

From the discontinuity in $E_r$ we have the expression for the surface
density of charge

\begin{eqnarray}\label{1}
4\pi \sigma = q a ^{-2} \sum\limits^{\infty}_{0} \left( 4n +1 \right)
\Phi _{2n} ~P_{2n}\left(\mu \right)~,
\end{eqnarray} 
so the condition that $q$ is the total charge gives $\Phi_0 =1$ on
integration over the sphere.  The discontinuity in $B_\theta$ gives
the surface current so

\begin{eqnarray} \label{2}
4\pi J_{\phi}=& \nonumber \\
& q\sin\theta a^{-2} \sum \frac{4n
+3}{(2n+2)(2n+1)} \chiup_{2n+1} P^{\prime}_{2n+1} \left(\mu\right)~,
\end{eqnarray}
where $P^{\prime}_{2n+1} = d/\!d\mu (P_{2n+1}).$

The condition that the current is solely due to the rotation of the
surface charge combines the above equations to give

\begin{eqnarray}\label{3}
 \sum^{\infty}_{n=0} \frac{4n+3}{(2n+2)(2n+1)}
\chiup_{2n+1}~P^{\prime}_{2n+1} = \nonumber \\
  \omega \sum^{\infty}_{0}
\left(4n+1\right) \Phi _{2n}~P_{2n}~.
\end{eqnarray}

Our final physical requirement is that at equilibrium there is no
surface component of the Lorentz force (because the surface is a
conductor)

\begin{equation*}
E_{\theta} + \omega \sin \theta B_r =0~,
\end{equation*}
which implies in terms of our potentials

\begin{eqnarray}\label{4}
\sum^{\infty}_0 \Phi _{2n} P^\prime_{2n}\left( \mu \right) =- \omega
\sum^{\infty}_0 \chiup_{2n+1} P_{2n+1} \left( \mu \right )~.
\end{eqnarray}

Our problem has now been reduced to the mathematical one of solving
equations (\ref{3}) and (\ref{4}) for the coefficients $\Phi_{2n}$ and
$\chiup_{2n+1}$ when $\omega \leqslant 1$ and is given, and $\Phi_0
=1$.

\section{\label{sec:mathsol}Mathematical Solution}

We need the following mathematical properties of the Legendre
Polynomials

\begin{equation*}
\int^{+1}_{-1} P^{\prime}_{2n+1}~P_{2m}~d \mu = 
\left\{
\begin{array}{lr}
2 & \qquad m\leqslant n\phantom{-1}\\
0 & \qquad m > n\phantom{-1}
\end{array}
\right.
\end{equation*}

\begin{equation*}
\int^{+1}_{-1} P^{\prime}_{2n}~P_{2m+1}~d \mu  =  
\left\{
\begin{array}{lr}
2 & \qquad m\leqslant n-1\\
0 & \qquad m > n-1 
\end{array}
\right.~,
\end{equation*}
together with their orthogonality relation.  Multiplying (\ref{4}) by
$\frac{1}{2} P_{2m+1}$ and integrating from $-1$ to $+1$ we find using
the above

\begin{eqnarray}\label{5}
\displaystyle\sum^{\infty}_{n=m+1} \Phi _{2n} = - \omega \eta_{2m+1}~,
\end{eqnarray}
where we have defined $\etaup_{2m+1}$ as $\chiup_{2m+1}/(4m+3)$
because it proves to be a more convenient variable to use.  Similarly
multiplying (\ref{3}) by $\frac{1}{2} P_{2m}$ and integrating we
find on writing the result in terms of $\eta$

\begin{eqnarray}\label{6}
\sum ^\infty _{n=m} \frac{(4n+3)^2}{(2n+2)(2n+1)} \etaup_{2n+1} = \omega~
\Phi_{2m}~.
\end{eqnarray} 

We shall give two different methods of solving (\ref{5}) and (\ref{6})
which are complementary in that one gives enlightenment on what is
happening in the other.

\subsection{\label{sec:firstmethod}First Method}

Subtracting from (\ref{5}) the same equation with $m+1$ written for
$m$ we obtain

\begin{eqnarray}\label{7}
\displaystyle\Phi_{2m+2} = \omega \left( \etaup_{2m+3}- \etaup_{2m+1}\right)~,
\end{eqnarray}
performing the same operation on (\ref{6}) and dividing by 4

\begin{equation}\label{8}
\frac{(m+\frac{3}{4})^2}{(m+1)(m+\frac{1}{2})}
\etaup_{2m+1} = -\frac{\omega}{4}(\Phi_{2m+2}-\Phi_{2m})~.
\end{equation}
Now the coefficient of $\etaup_{2m+1}$ is remarkably close to 1:
indeed it is 

\begin{equation}\label{9}  
\frac{m^2+\frac{3}{2}m+\frac{9}{16}}
{m^2+ \frac{3}{2} m+ \frac{1}{2}}
= 1+ \frac{1}{16(m+1)\left( m+ \frac{1}{2}\right)}
= 1+ \delta_m~,
\end{equation}
where
$\delta_0=\frac{1}{8},~\delta_1=\frac{1}{48},~\delta_2=\frac{1}{120},~
\delta_3=\frac{1}{224},~\delta_4=\frac{1}{360}$ etc, thus the
$\delta_m$ quite rapidly become small compared to one which suggests
that if only 1\% accuracy is needed we might neglect the $\delta_m$
for $m \geqslant 2$.  This will form the basis for a rapidly
convergent approximation scheme.  Using (\ref{7}) to eliminate the
$\Phi$ from (\ref{8}) we find for $m \geqslant 1$

\begin{equation*}  
\displaystyle\left( 1+ \delta_m \right) \etaup_{2m+1} = -
\frac{\omega^2}{4}\left(\etaup_{2m+3}- 2\etaup_{2m+1} + \etaup_{2m-1}
\right)~,
\end{equation*}
hence we obtain a recurrence relation with almost constant
coefficients

\begin{equation}\label{10}
\etaup_{2m+3} +2 \left[ 2\omega^{-2} \left( 1+\delta_m \right)-1
\right] \etaup_{2m+1} + \etaup_{2m-1} =0~.
\end{equation}

Thus $\etaup{_5} +2 \left[ 2\omega^{-2}\left(1+\frac{1}{48}\right) -1
\right] \etaup{_3} +\etaup{_1} =0$. Now, since $\omega \leqslant 1~,~
2\omega^{-2} -1 \geqslant 1$, so with a less than 2\% error in the
middle coefficient we can set $\delta_m =0$ for $m \geqslant 2~.$
(\ref{10}) is then reduced to a recurrence relationship with constant
coefficients.

The solution is

\begin{equation}\label{11}
\etaup_{2m+1}= C(-u)^m +D(-u)^{-m}   \qquad m \geqslant 2~,
\end{equation}
where $u$ and $u^{-1}$ are the roots for $t$ of

\begin{equation}\label{12}
t^2-2(2 \omega^{-2} -1) t +1 =0~,
\end{equation}
notice that

\begin{equation}\label{13}
\omega^2 = 4u/(1+u)^2 \ {\rm and}~ \sqrt{1-\omega^2} =
\sqrt{1-{\frac{v^2}{c^2}}} = \frac{1-u}{1+u}~.
\end{equation}

Evidently the product of the roots is one and the sum of the roots is
positive, so both roots are positive with $u<1,~ u^{-1}>1,~
{\rm{for}}\ \omega < 1$. However, the expression (\ref{11}) will
diverge oscillatorily due to the $u^{-1}$ term unless $ D=0$. Hence,

\begin{equation*}
\etaup_{2m+1} = C(-u)^m  \hspace{2cm}  m\geqslant 2~,
\end{equation*}
where,
\begin{equation}\label{14}
u=2 \omega^{-2} \left[ 1- \mbox{$\frac{1}{2}$} \omega^2
-\sqrt{1-\omega^2} \right]~.
\end{equation}
Notice that

\begin{equation*}
\omega \rightarrow 0,~u\rightarrow \mbox{$\frac{1}{4}$}
\omega^2~,
\end{equation*}
but as
\begin{equation*}
\omega \rightarrow 1,~u \rightarrow 1 -2 \sqrt{1-\omega^2}~.
\end{equation*}
In particular
\begin{equation}\label{15}
\etaup{_5} = Cu^2~,
\end{equation}
and
\begin{equation}\label{16}
\sum^{\infty}_2~ \etaup_{2m+1} = C \frac{u^2}{1+u}~.
\end{equation}

We still need to determine the constant $C$ as well as $\etaup{_3}$,~
$\etaup{_1}$ and the $\Phi_{2n}$.  Our strategy is to return to the
exact equations (\ref{6}) and (\ref{7}) for $m=0,1,2$ and to combine
them with the results (\ref{15}) and (\ref{16}) to determine the
unknowns $\Phi_2,~ \Phi_4,~ \etaup{_3},~ \etaup{_1}$ and $C$.

We rewrite (\ref{6})

\begin{equation}\label{17}
\sum^\infty_{n=m} \left( 1+ \delta_m \right) \etaup_{2n+1} =
\frac{\omega}{4} \Phi_{2m}~,
\end{equation}

\begin{equation}\label{18}
m=0 \ {\rm{gives\ }} \quad C \frac{u^2}{1+u} + \left( 1+\delta_1
\right) \etaup{_3} + \left( 1+\delta_0 \right) \etaup{_1} = \omega/4~,
\end{equation}

\begin{equation}\label{19}
m=1 \ {\rm{gives\ }} \quad C \frac{u^2}{1+u} + \left( 1+\delta_1
\right) \etaup{_3} = \frac{\omega}{4} \Phi_2~,
\end{equation}

\begin{equation}\label{20}
m=2 \ {\rm{gives\ }} \quad C \frac{u^2}{1+u}= \frac{\omega}{4}
\Phi_4~,
\end{equation}

\begin{equation*}
{\rm{with\ }}\qquad m=0\ (\ref{7}) \ {\rm{gives}} \quad \Phi_2= \omega
\left( \etaup{_3} - \etaup{_1} \right)~,
\end{equation*}

\begin{equation*}
{\rm{with\ }}\qquad m=1\ (\ref{7})\ {\rm{gives}} \quad \Phi_4= \omega
\left( Cu^2- \eta_3 \right)~.
\end{equation*}
So subtracting (\ref{19}) from (\ref{18})

\begin{equation}\label{21}
\left( 1+ \delta_0 \right) \etaup{_1} = \frac{\omega}{4} \left[ 1-
\omega \left( \etaup_3 - \etaup{_1} \right) \right]~,
\end{equation}
while (\ref{19}) and (\ref{20}) become
\vspace{-.4cm}

\begin{equation}\label{22}
C \frac{u^2}{1+u}+ \left(1+\delta_1 - \frac{\omega^2}{4}\right) \etaup{_3}
+ \frac{\omega^2}{4}\etaup{_1} =0~,
\end{equation}

\begin{equation*}
{\rm and}\qquad  C \left(\frac{u^2}{1+u} - \frac{\omega^2}{4} u^2
\right)= - \frac{\omega ^2}{4} \etaup{_3}~,
\end{equation*}
using (\ref{13}) to re-express $\frac{\omega^2}{4}$ as $\frac{u}{(1+u)^2}$
we find 

\begin{equation}\label{23}
\etaup_3 =-Cu~,
\end{equation}
putting (\ref{23}) into (\ref{22}) we find

\begin{eqnarray}\label{24}
\etaup_1 = \left[ 1+\delta_1 (1+u)^2 \right] C~,
\end{eqnarray}
finally putting expressions (\ref{23}) and (\ref{24}) for $\etaup_3\
{\rm{and}\ } \etaup_1$ into ({\ref{21}) we have our equation for the
constant $C$ again using (\ref{13})

\begin{equation}\label{25}
C= \frac{1}{2} u^\frac{1}{2} / \Delta~,
\end{equation}
where

\begin{equation}\label{26}
 \Delta =1 + \left( 1+u \right) \left[ \delta_0 + \delta_1 \left(1+u +
 u^2 \right) + \delta_0~ \delta_1 \left( 1+u \right )^2 \right]~,
\end{equation}
with $\delta_0 = \frac{1}{8}$ and $ \delta_1 = \frac{1}{48}$ and $u$
given by (\ref{14}).

With $C$ so determined $\etaup_1$ is given by (\ref{24}) and
$\etaup_3$ by (\ref{23}) may be incorporated into the more general
formula

\begin{equation*}
\etaup_{2m+1} = C(-u)^m \ \qquad {\rm{for\ }}m \geqslant 1~.
\end{equation*}

For $u<0.9$ a somewhat more accurate solution may be shown to be

\begin{equation*}
\etaup_{2m+1}= C(-u)^m \left[1+\frac{(1+u)}{16(1-u)}~
\frac{1}{\left(m+\frac{7}{4}\right)}\right]~.
\end{equation*}

Its accuracy is $1.5\%$ at $u=.8$ but diminishes to 7\% at $u=.9$.
Beyond that we give a separate treatment in section 3.2.

With all the $\eta$ known the $\Phi_{2m}$ are given by (\ref{7})
with $\omega$ given by (\ref{13}).  Thus 

\begin{eqnarray*}
\Phi_0 & = & 1~, \nonumber\\
& & \nonumber\\ 
\Phi_2 & = & - u\left[1+ \delta_1 (1+u) \right] / \Delta~, \nonumber\\
& & \nonumber\\
\Phi_{2m}& = & (-u)^m/ \Delta \hspace{2.5cm} m\geqslant 2~.
\end{eqnarray*}
Hence

\begin{eqnarray}\label{27}
\Sigma &=& \sum^{\infty}_1 \Phi^2_{2m} \nonumber \\
&=& \frac{u^2}{\Delta^2} \left[
  \frac{1}{(1-u^2)} +2\delta_1(1+u) + \delta^2_1 (1+u)^2 \right]~.
\end{eqnarray}

Those who wish for a much more accurate calculation including the
first order perturbation theory for the terms neglected above, will
find it in the appendix.  However, there is always some tension
between simple explanation and a detailed calculation in which the
reader may get lost in the mass of formulae, so I have chosen the path
of simplicity for the main paper which helps to keep the physics to
the fore and to suppress the minutiae of mathematical manipulation.
The above approximate solutions for the potentials are
\begin{equation} \label{28}
\Phi = \left\{
\begin{array}{lll}
\frac{q}{r}\left(1-\frac{1}{\Delta}\right)
+\frac{q}{\Delta}\left[\frac{1}{r}\sum\limits^\infty_{n=0}
\left(-u\frac{a^2}{r^2}\right)^nP_{2n}(\mu) \right. 
\\ 
\left. -\delta_1u(1+u)\frac{a^2}{r^3}P_2(\mu)\right]~,\qquad\quad r\geqslant a~.
 \\ \frac{q}{a}\left(1-\frac{1}{\Delta}\right)
+\frac{q}{a\Delta}\left[\sum\limits^\infty_{n=0}
\left(-u\frac{r^2}{a^2}\right)^nP_{2n}(\mu) \right. \\
\left. -\delta_1u(1+u)\frac{r^2}{a^2}P_2(\mu)\right]~,\qquad\quad r\leqslant a~.
\end{array}
\right.
\end{equation}
Now

\begin{equation}\label{29}
\left(r^2-2 \mu br+b^2 \right)^{-\frac{1}{2}} = \frac{1}{r}
 \sum \left( \frac{b}{r}\right )^n ~ P_n (\mu)~, \qquad r\geqslant |b|~.
\end{equation}
So if we put $b = ia \sqrt{u} $, the real part is

\begin{equation*}
\frac{1}{r} \sum^\infty_0 \left(-u \right)^n \left( \frac{a^2}{r^2}
  \right)^n P_{2n}(\mu) \qquad r \geqslant a~,
\end{equation*}
which should be compared to the first of the sums in $\Phi$.

We see that our potential is that due to charges $\frac{1}{2}  q/
\Delta$ at the imaginary points $z = \pm~ ia \sqrt{u}$ plus a quadrupolar
correction term in $\delta_1P_2$ and a charge $q(1-\frac{1}{\Delta})$
at the origin.  Whereas that construction gives us the field outside
the sphere, the field inside the sphere can be thought of as arising
from charges at the imaginary points conjugate to them under inversion
in the sphere $r=a$ plus a quadrupolar correction term in $\delta_1 P_2$.

The magnetostatic potential $\chiup$ takes the form for $r \geqslant
a$

\begin{equation*}
\chiup = 
\frac{qu^{\frac{1}{2}}}{\Delta} 
\left \{ \sum_n
\frac{2n+\frac{3}{2}}{2n+2} \left( 
\frac{a}{r} \right)^{2n+2} (-u)^n +
\frac{3}{4} \delta_1 (1+u)^2 
\frac{a^2}{r^2} \mu \right\}~.
\end{equation*}

\medskip

We notice that putting $b=i\sqrt{u}a$ and taking the imaginary
part of (\ref{29}) yields for $r \geqslant a$

\begin{equation*}
\mathcal{I}m \left( r^2 -2i \mu \sqrt{u}~a-a^2 u \right)^{-\frac{1}{2}} = 
\frac{u^{\frac{1}{2}}}{a} \sum \left( -u \right)^n \left( \frac{a}{r}
\right)^{2n+2}~,
\end{equation*}
also
\begin{eqnarray*}
\mathcal{I}m \int^a_0 \left(  r^2-2i \mu \sqrt{u}~ ar - a^2u
\right)^{-\frac{1}{2}} a^{-1} da = \\
\frac{u^\frac{1}{2}}{a} \sum
\frac{(-u)^n}{2n+1} \left(\frac{a}{r}\right)^{2n+2}~.
\end{eqnarray*}
Hence at least for $r \geqslant a,~ \chiup$ is of the form 

\begin{eqnarray}
\chiup &=& \frac{q}{\Delta} \mathcal{I}m \left [ \left( r^2 - 2i \mu
    \sqrt{u}~ ar - ua^2 \right)^{-\frac{1}{2}} - \right. \nonumber \\
& -& \left. \frac{1}{2} \int^a_0
    \left( r^2 - 2i \mu \sqrt{u}~ ar-ua^2 \right) ^{-\frac{1}{2}}
    \frac{da}{a} \right ]\nonumber \\
& +& \frac{3 \delta_1(1+u)^2}{4 \Delta} q \frac{au^
    \frac{1}{2}}{r^2}~\mu,  \label{30}
\end{eqnarray}
within the square bracket the first term is clearly the magnetic field
due to the charge $\frac{q}{\Delta}$ at the imaginary point $i
\sqrt{u}~a \mathbf{\hat{z}}$ with the convention that $\mathbf{E} + i
\mathbf{B} = - \nabla \left[ \Phi + i \chiup \right]$ this fits nicely
with our result for the potential.  The second term in the square
bracket may be interpreted at the magnetic potential of the charge
distribution $\frac{q}{\Delta}~ \frac{da^\prime}{a^\prime}$ at $i
\sqrt{u}~ a^\prime \mathbf{\hat{z}}$ between $a^\prime = 0$ and
$a^\prime=a$.  However, there is no corresponding part in the
electrical potential.

\subsection{\label{sec:limitv=c}The Limit $v=c$}

When $v/c$ is within half of a percent of 1 (ie $\gamma > 20$) even
the amazingly good perturbation theory of the Appendix which allows
linearly for all the terms we have neglected, breaks down; it generates
terms in $\frac{1}{8} \ln (1-u)$ which can no longer be treated as
perturbations; we therefore need a new approach for that very small
region.  Our exact difference equation for $\etaup_{2n+1}$ is

\begin{eqnarray*}
\etaup_{2n+3} + \left[u+\frac{1}{u} + \frac{(u+1)^2}{16u}
\frac{1}{\left(n+\frac{1}{2}\right )\left(n+1 \right)} \right]
\etaup_{2n+1} \\
+ \etaup_{2n-1} =0~,
\end{eqnarray*}
at $u=1$ the solution when the $\frac{1}{n^2}$ term is neglected is
$\etaup_{2n+1}= C(-1)^n$.  To solve the equation without any such
neglect, we write $\etaup_{2n+1}= (-1)^n v_n$ and obtain

\begin{eqnarray} \label{31}
\left( v_{n+1}-v_n \right)-\left( v_n -v_{n-1}\right)- \nonumber\\
\left[
    \frac{(u-1)^2}{u}+ \frac{(u+1)^2}{16u}
    \frac{1}{\left(n+\frac{1}{2} \right) \left(n+1 \right)} \right]
    v_n =0~.
\end{eqnarray}

Let us look for solutions in which $v_n$ varies sufficiently slowly
with $n$ that we may approximate $v_n$ as a continuous function
$v(n)$.  Then $(v_{n+1}-v_n)-(v_n-v_{n-1}) = \frac{d^2v}{dn^2}$.

Thus to sufficient accuracy we have for large $n$
\begin{equation} \label{32} 
d^2 v/dn^2 = \left[\frac{(u-1)^2}{u}+\frac{(u+1)^2}{16u}
\frac{1}{\left(n+\frac{3}{4}\right)^2} \right] v(n)~.
\end{equation}

The solution that converges at large $n$ is a Bessel function of
imaginary argument. Setting $N=n+\frac{3}{4}$

\begin{equation*}
v(n)=C \sqrt{\frac{2}{\pi}} \left( \frac{1-u}{u^\frac{1}{2}} N
\right)^\frac{1}{2} K_\nu \left( \frac{1-u}{u^\frac{1}{2}} N \right)~,
\end{equation*}
where

\begin{eqnarray*}
\nu^2=\frac{1}{4}\left(1+\frac{1}{4}~\frac{(u+1)^2}{u}\right)\\ v(n)
\rightarrow C \pi^{-\frac{1}{2}} ~\Gamma (\nu)
\left(\frac{1-u}{2u^\frac{1}{2}}N \right) ^{\frac{1}{2} - \nu}~,
\end{eqnarray*}
for small $\frac{1-u}{\sqrt{u}}N$ and to
$C\exp\left(-\frac{1-u}{\sqrt{u}}N\right)$ for large.

In the relevant r\'{e}gime with $u \sim 1$ then $\nu =\frac{1}{\sqrt{2}}$~.

It is more helpful to derive the correct form for $n$ large, as we
shall do presently.  However, before that we should assure ourselves
that our technique works by applying it to the equation in which the
$\left(n+\frac{3}{4}\right)^{-2}$ terms are absent.  Then the exact
convergent solution is $v_n=Cu^n$ and near $u=1$ this may be
re-expressed as

\begin{eqnarray*}
v_n=C\exp \left\{n\ln \left[1-(1-u) \right] \right\} = \\
C \exp \left[
  -n (1-u)   \right] + 0(1-u)^2 ~.
\end{eqnarray*}

Equation (\ref{32}) gives a solution 

\begin{equation*}
v(n)=C \exp \left[ -n (1-u) u ^{-\frac{1}{2}} \right ]~,
\end{equation*}
which is the same to order $(1-u)^2$.  Thus our approximation only
works in the neighbourhood of $u=1$ but that is where we need it.
When $u=1$ equation (\ref{32}) has the exact solution $v(n)=C_1 \left( n+
\frac{3}{4} \right)^{-\alpha}$ where $\alpha = 1/ \left( 2 \sqrt{2} +2
\right ) \simeq 0.207107$ so this must be the asymptotic form of the
solution of the difference equation for $v_n$ when $n$ is large.

Writing $N=n+\frac{3}{4}$ the exact equation at $u=1$ is 

\begin{equation*}
v_{n+1} + v_{n-1} = \left[2 + \frac{1}{4 \left( N^2- \frac{1}{16}\right)}
\right ] v_n~,
\end{equation*}

\begin{equation*}
\left[
\begin{array}{l}
 {\rm to\ be\ exact\ away\ from\ } u=1 \ {\rm an\ extra\ term\ }\\ +
\frac{(u-1)^2}{u}\left[1+ \frac{1}{16 \left ( N\frac{2}{}
\frac{1}{16}\right)} \right] v_n \ {\rm is\ needed\ on\ the\ right }
\end{array}
\right]~,
\end{equation*}

We look for solutions of the form 

\begin{equation*}
v_n = C_1 N^{- \alpha} \left( 1+AN^{-2} +BN^{-4} \right)~.
\end{equation*}

Terms of order $N^{-2-\alpha}$ in the recurrence relation give us the
value of $\alpha$ found above and the terms of order $N^{-4-\alpha}$
can be used to give us an equation for $A$.  This is 

\begin{equation*}
A= -\frac{1}{96} \left[ 2 - \frac{1}{2 (2 \alpha +3)} \right ] = -
0.0193078~.
\end{equation*}

We chose the form with only even powers, as then the odd powers
cancel.  The next terms are of order $N^{-6}$ times the leading terms.
These give 

\begin{equation*}
\begin{array}{rcl}
\left( 4 \alpha + 10 \right ) B & = & - \frac{1}{24} \left( 130 \alpha
  + 134 \frac{7}{16} \right) \left(\frac{1}{120} + A \right) + \\ 
&&\\
& + &
  \frac{1}{2048} + \frac{A}{120}\\ 
&& \\
\raisebox{0.2em}{$\therefore$}~ B &
  = & 0.0068443~.
\end{array}
\end{equation*}
We have checked by computer that the asymptotic formula with these
values of $A$ and $B$ works down to $n=0$ with an error with less than
one part in $1000$, so

\begin{equation} \label{33}
\etaup_{2n+1} = C_1 (-1)^n N^{- \alpha} \left[1+AN^{-2} + BN^{-4}
\right]
\end{equation}
where $N=n+\frac{3}{4}$ is the accurate convergent solution with the
above values of $\alpha,~ A,~ B$.  The constant $C_1$ is determined
via equation (\ref{6}) with $m=0~,~ \Phi_0 =1~,~ \omega = 1$

\begin{eqnarray} \label{34}
C_1 &=& \frac{1}{4}\Bigg/ \sum^{\infty} _{n=0}\frac{(-1)^n
    N^{-\alpha}}{1- \left( \frac{1}{16 N^2}\right)} \left[1+
    AN^{-2} + BN^{-4} \right]  \nonumber \\
&\bumpeq&  \frac{1/4}{0.699}= .358~.
\end{eqnarray}

The convergence of the sum is slow $2 \times 10^9$ terms gave $.6926$
but comparison with
$2\times\left(10^1,~10^2,~10^3,~10^4,~10^5\right.$,
$\left.10^6,~10^7,~10^8 \right)$ terms show the increments decrease in
geometrical progression by a factor $1.608$, so extrapolation is
accurate.

In this r\'{e}gime we can now calculate $\etaup_1 = .375 = \frac{3}{8}$.

\section{\label{sec:explicit}Explicit Expressions for Potentials, 
Charges and Currents}
 
Using (\ref{29}) we write expression (\ref{28}) for $\Phi$ in the form

\begin{equation} \label{35}
\Phi = \left\{
\begin{array}{lll}
\frac{q}{r} \left[1 - \frac{1}{\Delta} \left(1
      +\frac{a^2}{r^2}\frac{(1+u)u}{48} P_2 \left(\mu \right)
      \right)\right ]\\ 
+q \frac{1}{\Delta} Re \left( r^2-
      2iu^\frac{1}{2} ar \mu - ua^2 \right)^{-\frac{1}{2}}~, \qquad r
      \geqslant a \\ 
& & \\ \frac{q}{a} \left[1 -
      \frac{1}{\Delta} \left(1 +\frac{r^2}{a^2}\frac{(1+u)u}{48} P_2
      \left(\mu \right) \right)\right ]\\ 
+ \frac{q}{\Delta} Re \left(
      a^2+2iu+^{\frac{1}{2}} ar \mu - ur^2 \right)^{-\frac{1}{2}}~,
      \qquad r \leqslant a~.
\end{array}
\right.
\end{equation}

Notice that the points $\left(0,~0,~iu ^\frac{1}{2} a \right )$ and $
\left( 0,~0,~-iu^{-\frac{1}{2}} a \right )$ are inverse points for the
sphere.  The surface density of charge is given in terms of the
discontinuity in $\partial \Phi / \partial r {\rm{\ at\ }} r=a$~.

\begin{eqnarray*}
4 \pi \sigma = q a ^{-2} \left\{1- \frac{1}{\Delta} \left ( 1 +
\frac{5}{48} u (1+u) P_2 \right )+ \right. \\
+ \left. \frac{1}{\Delta} (1+u) Re \left(
1+2iu^\frac{1}{2} \mu -u \right) ^{-\frac{3}{2}} \right \}~.
\end{eqnarray*}

Putting $\mu = 0$ we obtain the values on the equator

\begin{eqnarray} \label{36}
\sigma _{\rm{equ}} = \frac{q}{4 \pi a ^2} \left [ 1- \frac{1}{\Delta}
  \left( 1-\frac{5}{96} u (1+u) \right ) \right. + \nonumber\\
+ \left. \frac{1+u}{\Delta(1-u)^\frac{3}{2}} \right ] ~,
\end{eqnarray}
this clearly diverges as $u$ approaches $1$ because $\Delta$ remains
finite.  Similarly putting $\mu =1$ and evaluating $\left (
1+iu^\frac{1}{2} \right )^{-3}$

\begin{eqnarray} \label{37}
\sigma_{\rm{pole}} = \frac{q}{4 \pi a^2} \left[1- \frac{1}{\Delta}
\left( 1+ \frac{5}{48}u (1+u) \right ) + \right. \nonumber\\
\left.\frac{1-3u}{\Delta(1+u)^2}
\right] ~.
\end{eqnarray}

This density changes sign as $u$ increases.  Taking account of the
variation of $\Delta$ given in (\ref{26}), we find
$\sigma_{\rm{pole}}=0$ at $u= 0.454,~ \gamma = 2.66,~ \omega = 0.93~.$
The surface current is solely due to the advection of the surface
charge so $J _{\phi} = \sigma \omega \sin \theta$.

We may get a good idea of the value of $\theta$ or ($\mu$) at which
the charge density changes sign as follows; $\frac{1}{\Delta}$ is not
far from $1$.  In fact $1- \frac{1}{\Delta} = \frac{f}{1+f}$ where by
(\ref{26})

\begin{equation*}
f=\frac{1}{8} (1+u) \left[ 1+ \frac{1}{6} \left( 1+u +u^2 \right) +
\frac{1}{48} (1+u)^2 \right]~,
\end{equation*}
so, $\frac{f}{1+f}$ varies between $\frac{1}{9}$ and $\frac{19}{67}$~.
The major variation $\sigma$ with $\mu$ is caused by

\begin{equation*}
\left( \frac{1+u}{\Delta} \right) Re \left( 1+2iu ^\frac{1}{2} \mu - u
\right )^{-\frac{3}{2}}~,
\end{equation*}
which becomes $>6 {\rm{\ for\ }} (1-u)< \frac{1}{3}$.  We write 

\begin{equation*}
\left( 1+2iu ^\frac{1}{2} \mu -u \right) = \qquad Me^{i \beta}~,
\end{equation*}
where

\begin{equation*}
\beta = \tan^{-1} \left( {2u^\frac{1}{2} \frac{\mu}{1-u}} \right)~,
\end{equation*}
and
 
\begin{equation*}
M^2 = (1-u)^2 + 4u \mu ^2~.
\end{equation*}

Evidently $M^{-\frac{3}{2}} \cos \left( \frac{3}{2} \beta \right)$
will become zero when $\beta = \frac{2}{3} \frac{\pi}{2}$ i.e. where
$2u^\frac{1}{2} \mu / (1-u) = \tan \frac{\pi}{3} = \sqrt{3}$, so this
corresponds to $\mu = \frac{1}{2} \sqrt{\frac{3}{u}} (1-u)$ which
becomes small when $u$ approaches one.  Thus the charge of the same
sign as $q$ becomes confined to a belt between latitudes

\begin{equation*}
\pm \sin ^{-1} \left[ \frac{1}{2} \sqrt{\frac{3}{u}} (1-u) \right ]
\rightarrow \sqrt{3} \sqrt{1-\frac{v^2}{c^2}}~,
\end{equation*}
and the total width of the belt around the equator $\rightarrow
\sqrt{3} a(1-u)$.

By then most of the sphere's surface is covered with charge of the
opposite sign leaving only the equatorial belt with a much enhanced
charge of the same sign as $q$.  The charge in that belt is
approximately

\begin{equation*}
\frac{q (1+u)}{2 \sqrt{2} \Delta (1-u) ^\frac{1}{2}}
\end{equation*}
which tends to $\infty$ like $\frac{q}{2 \Delta} \left ( 1-
\frac{v^2}{c^2} \right)^{-\frac{1}{4}}$ but these only hold when $v$
is close to $c$.  A rough approximation to the charge in the belt more
generally is

\begin{equation*}
q \left[ 1+ \frac{1}{2 \Delta} \left( \gamma^\frac{1}{2} -
\gamma^\frac{1}{2}_0 \right)U \left( \gamma - \gamma_0 \right)
\right]~,
\end{equation*}
where $U$ is one for positive values and zero otherwise and $\gamma_0$
is the value of $\gamma$ at which the charge changes sign at the pole
ie $\gamma_0 = \frac{1.454}{.546} = 2.66$~.

Although all our calculations thus far have been in terms of the
magnetostatic potential $\chiup$ we shall find that several global
properties of the electromagnetic field are more nicely expressed in
terms of the vector potential $\mathbf{A}$.  Since $\mathbf{B}$ lies
in meridional planes and is axially symmetrical, we may take
$\mathbf{A}$ in the form $- \zeta~ \nabla~ \phi $ where $\phi$ is the
azimuthal angle of spherical polar co-ordinates.  Then

\begin{equation*}
\begin{array}{lcl}
\displaystyle \mathbf{B} & = & {\rm{Curl}}\ \mathbf{A}= - \nabla \zeta
\times \nabla \phi\\ & &\\ \displaystyle B_r & = &- \partial \chiup /
\partial r = - \frac{1}{r^2 \sin \theta} \partial \zeta/ \partial
\theta = + r^{-2} \partial \zeta/ \partial \mu~.
\end{array}
\end{equation*}

Inserting our expansion for $\chiup$ in terms of the $P_{2n+1}$

\begin{equation*}
\frac{q}{2a^2} \sum \left(\frac{a}{r} \right)^{2n+3} \chiup_{2n+1}~ 
P_{2n+1} = + r^{-2}~ \frac{\partial \zeta}{\partial \mu}~,
\end{equation*}
but the $P_n$ obey Legendre's equation

\begin{equation*}
- \frac{1}{(2n+1)(2n+2)} \frac{d}{d \mu} \left[ \left( 1- \mu^2
  \right) \frac{dP_{2n+1}}{d \mu} \right ] = P_{2n+1}~,
\end{equation*}
so inserting this into the expression above and using the boundary
condition that $\zeta = 0$ when $ \mu =+1$ we find

\begin{equation*}
q \sum \left( 
\frac{a}{r} \right)^{2n+1}
\frac{\chiup_{2n+1}}{(2n+1)(2n+2)} ~(1-\mu^2) \frac{dP_{2n+1}}{d \mu}
=-\zeta
\end{equation*}
for $r\geqslant a$ and a very similar condition for $r\leqslant a$.

\begin{figure}
\includegraphics{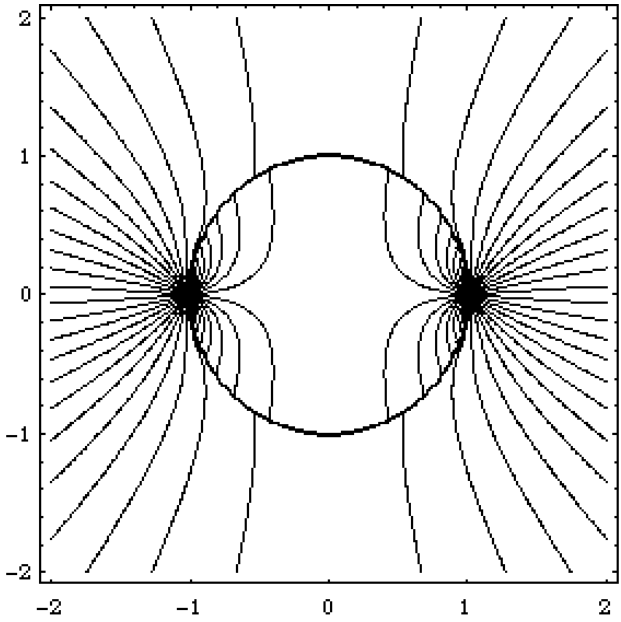}
\caption{\label{fig:figure} The electric lines of force for a
relativistically rotating conducting spherical shell with u=0.9
i.e. gamma=19, v/c=0.9986.  Notice the charge concentration at the
equator and the diminution of the field strength as the lines cross
the sphere again, due to the opposing charge there. In contrast the
static shell has a radial field outside and none inside.}
\end{figure}

Now 

\begin{equation*}
(1-\mu^2)dP_n/d\mu = n(P_{n-1}-\mu P_n)~,
\end{equation*}
and 
\begin{equation*}
\mu P_n =
[(n+1)P_{n+1}+n P_{n-1}]/(2n+1)~,
\end{equation*}
so 
\begin{equation*}
(1-\mu^2)dP_n/d\mu =
\frac{n(n+1)}{2n+1}(P_{n-1}-P_{n+1})
\end{equation*}
thus
\begin{equation*}
q\sum\left(\frac{a}{r}\right)^{2m+1}\frac{\chiup_{2n+1}}{4n+3}~\left(P_{2n+2}
-P_{2n}\right) = \zeta
\end{equation*}
Notice that $\chiup_{2n+1}/(4n+3)$ is precisely our $\etaup_{2n+1}$ so 

\begin{equation}\label{38}
\zeta=-q \left\{
\begin{array}{ll}
\sum\left(\frac{a}{r}\right)^{2n+1}\etaup_{2n+1}(P_{2n}-P_{2n+2})
~ r\geqslant a \\
& \\
\sum\left(\frac{r}{a}\right)^{2n+1}\etaup_{2n+1}(P_{2n}-P_{2n+2})
~r\leqslant a 
\end{array} \right.
\end{equation}
A similar change of potential shows that the electric field can be
written as $\nabla\phi \times\nabla \bar{\zeta}$ where

\begin{eqnarray}\label{39}
\bar{\zeta}=q
\left[\mu\left(1-\frac{1}{\Delta}\right)+\frac{a^2u(1+u)}{\Delta
r^2~32}\mu\left(1-\mu^2\right)\right]
+ \nonumber \\ 
\frac{q}{\Delta}Re\left\{\frac{z-iau^{1/2}}{[({\bf r}
-i{\bf a}\sqrt{u})^2]^{1/2}}\right\} \hspace{1cm} r\geqslant a
\end{eqnarray}
The above expression is useful as the contours of constant
$\bar{\zeta}$ give the electric lines of force. Whereas these are
radial for a non-rotating sphere they are more interesting for high
rotation speeds and are plotted for $u=0.9$, $\gamma=19$ as Figure 1.

\section{\label{sec:global}Global properties of the Electromagnetic Field}

The Poynting vector is $\frac{c}{4\pi}{\bf E}\times{\bf B}$ and as this
flows around the axis it gives an angular momentum density

\begin{equation*}
\frac{1}{4\pi c}~{\bf r} \times ({\bf E} \times {\bf B})~.
\end{equation*}
Thus the total angular momentum in the electromagnetic field is

\begin{eqnarray*}
{\bf L}=\frac{1}{4\pi c}\int{\bf r}\times ({\bf E}
\times {\bf B})d^3x = +\frac{c}{4\pi}\int
{\bf r} \\
\times[\nabla\Phi\times (\nabla
\zeta\times \nabla \phi)]d^3x~,
\end{eqnarray*}
where we have used axial symmetry and the fact that ${\bf B}$ lies in
meridional planes to write the vector potential in the form
$-\zeta\nabla\phi$ where $\phi$ is the spherical polar coordinate.

Since $\nabla^2\Phi=0$ we deduce that 

\begin{equation*}
\mathbf{L}=\frac{-1}{4\pi c}\int{\mathbf{r}}\times
\nabla\phi \nabla\Phi . \nabla\zeta
d^3r= \frac{-\hat{\mathbf{z}}}{4\pi c}\int\nabla\Phi . \nabla\zeta d^3r~,
\end{equation*}
where $\mathbf{R}\times\nabla\phi=\hat{\bf z}$ and
$\mathbf{R}=(x,y,0)$ radial components of $\mathbf{r}\times\nabla\phi$
average to zero around a circle about the axis, by symmetry.

Now $\nabla\Phi . \nabla\zeta={\rm
div}(\zeta\nabla\Phi)-\zeta\nabla^2\Phi$ and $\nabla^2\Phi$ is zero
both within and outside the sphere. Hence the angular momentum in the
field within the sphere is

\begin{eqnarray*}
{\bf L}_i&=&\frac{-\hat{\mathbf{z}}}{4\pi c}
\int{\rm div}(\zeta\nabla\Phi)d^3r 
= \frac{-\hat{\mathbf{z}}}{4\pi c}
\int \zeta\nabla\Phi . d^2S~, \\
&=&\frac{-\hat{\mathbf{z}}}{2c}a^2
\int^{+1}_{-1}\zeta
\frac{\partial \Phi}{\partial r}\bigg|_{a-}a^2d\mu~,
\end{eqnarray*}
where the gradient $\partial\Phi/\partial r$ is evaluated on the inside
of $r=a$. Similarly for the angular momentum external to the sphere

\begin{equation*}
{\bf L}_o=+\frac{\hat{\mathbf{z}}a^2}{2c}\int
\zeta\frac{\partial\Phi}{\partial r}\bigg|_{a+}d\mu~,
\end{equation*}
so ~the ~total ~angular ~momentum ~depends ~on
\newline
$\left(\frac{\partial\Phi}{\partial
r}\right)_{a+}-\left(\frac{\partial\Phi}{\partial r}\right)_{a-}=-4\pi
\sigma$ and we find

\begin{equation*}
{\bf L} = {\bf L}_o+{\bf L}_i=-\hat{\bf z}2\pi a^2
c^{-1}\int^{+1}_{-1}\zeta\sigma d\mu~.
\end{equation*}
Inserting the expressions for $\zeta$ and $\sigma$ from the last section

\begin{eqnarray*}
{\bf L}=\frac{q^2\hat{\mathbf{z}}}{2c}\int^{+1}_{-1}
\left[\sum^\infty_{n=0}\etaup_{2n+1}(P_{2n}-P_{2n+2})\right] \times \\
\times \sum^\infty_{m=0}(4m+1)\Phi_{2m}P_{2m}d\mu~.
\end{eqnarray*}
Most of the terms vanish due to the orthogonality of the $P_n(\mu)$
and we are left with (defining $\etaup_{-1}\equiv 0)$

\begin{equation*}
{\bf L}=\hat{\mathbf{z}}\frac{q^2}{c} \sum^\infty_0
\left[(\etaup_{2m+1}-\etaup_{2m-1})\Phi_{2m}\right]~.
\end{equation*}
But by equation (7) this is just (remembering $\Phi_0=1)$

\begin{equation}\label{40}
{\bf L}=\hat{\mathbf{z}}\frac{q^2}{c\omega}\left(\Sigma +
\omega~\eta_1\right)~~~~{\rm where~} \Sigma=\sum^\infty_1\Phi^2_{2m}~.
\end{equation}
These formulae are exact without any approximation. In a rather
similar manner one may derive exact formulae for the electical energy
$\epsilon_e=\int (E^2/8\pi)d^3r$ and the total both inside and out gives

\begin{equation}\label{41}
\epsilon_e=\frac{1}{8\pi} \int^{+1}_{-1}\Phi 4\pi\sigma 2\pi a^2 d\mu=
\frac{q^2}{2a} \sum^\infty_0\Phi^2_{2n} =
\frac{q^2}{2a}\left(1+\Sigma\right)
\end{equation}
For the magnetic energy we have
\begin{eqnarray}\label{42}
\epsilon_m&=& \frac{1}{8\pi}\int ({\rm
curl}{\bf A})^2d^3r =\frac{1}{8\pi} \int {\rm div}({\bf A} \times {\bf
B})d^3r \nonumber \\ 
&=&\frac{1}{8\pi} \left[\int({\bf A} \times {\bf
B})_idS - \int({\bf A}\times {\bf B})_0dS\right] \nonumber\\
&=& \frac{1}{8\pi}\int
A_\phi [{\bf B}_o-{\bf B}_i]_\theta 2\pi a^2 d\mu \nonumber \\
&=&\frac{a^2}{4}\int A_\phi 4\pi J_\phi d\mu = -\frac{a^2}{4}
\int\frac{\zeta}{a\sin\theta}4\pi \sigma \omega\sin\theta d\mu \nonumber \\
&=&\frac{-\omega q}{4a}\int^{+1}_{-1} \sum_m (4m+1)\Phi_{2m}P_{2m}d\mu
\nonumber \\ 
&=&\frac{1}{2} \frac{q^2}{a} \left(\Sigma +
\omega\eta_1\right) =\frac{1}{2}\frac{Lc\omega}{a}
\end{eqnarray}
Thus
\begin{equation}\label{43}
\epsilon =\epsilon_e+\epsilon_m=\frac{q^2}{a}\left[\frac{1}{2}+\Sigma 
+\frac{1}{2}\omega\eta_1\right]~.
\end{equation}
$\Sigma$ is given by (27) and $\etaup_1$ by (24) in our first
approximation.

\section{\label{sec:dipolequad}The Dipole and Quadrupole Moments}

Directly from our expressions for $\Phi$ and $\chi$ as series we may
read off the dipole and quadrupole moments of the field. There is no
electric dipole and the magnetic dipole is

\begin{equation}\label{44}
\mu_m=\frac{1}{2}q a\chiup_1= \frac{3}{2}qa\eta_1
\end{equation}
similarly there is no magnetic quadrupole moment but the electric
quadrupole is

\begin{equation*}
Q_e=qa^2\Phi_2
\end{equation*}
similarly the odd moments are missing from electric field and the even
moments are missing from the magnetic field. We shall not detail the
$(2n+1)^{\rm th}$ order moments but we note that as $v\rightarrow c$
they only fall as $(n+3/4)^{-\alpha}$ with $\alpha$ close to
$0.2$. This slow fall leads to a divergent total field energy as
$v\rightarrow c$. This may be seen by comparing the series in the sum
$\Sigma$ with that for Riemann's zeta function $\zeta(1)$ which is
infinite.

The gyromagnetic ratio is $\mu_m/L$ and the g-factor
$\frac{2mc}{q}~\frac{\mu_m}{L}$ so

\begin{equation*}
g=\frac{3\etaup_1}{(\epsilon/mc^2)}~~\frac{(\frac{1}{2}+\Sigma+
\frac{u^{1/2}}{1+u}\etaup_1)}{\left(\etaup_1+\frac{1+u}{2u^{1/2}}\Sigma\right)}
\end{equation*}
This formula along with (\ref{40}) assumes that all the angular momentum is
in the moment of the Poynting vector of the electromagnetic
field. When the system is highly relativistic $\sum$ dominates so

\begin{equation*}
g\rightarrow \frac{2\etaup_1}{(\epsilon/mc^2)}\simeq
\frac{9}{8}\left(\frac{mc^2}{\epsilon}\right)
\end{equation*}
where $\epsilon$ is the energy in the field. Notice that the presence
of non-electromagnetic mass will {\bf increase} g above 9/8 if we have
only electromagnetic angular momentum.

\section{\label{sec:oldelectron}Old Fashioned Electron Models}

A century ago it would have been natural to apply the foregoing
results to Mie's `model' of the electron. Now everyone knows better
(and some like me understand less). However in the spirit of those
times let us set the charge equal to $e$, the dipole moment equal to
$e\hbar/2mc$ and the total field angular momentum equal to
$\frac{1}{2}\hbar$. Then

\begin{equation}\label{45}
\mu_m=\frac{e\hbar}{2mc} =\frac{3}{2}ea~\eta_1
\end{equation}
\begin{equation}\label{46}
\frac{Lc}{e^2}=\frac{\hbar c}{2e^2}= \etaup_1 +\frac{1+u}{2u^{1/2}}\Sigma
\end{equation}
and the total electromagnetic energy is then

\begin{eqnarray}\label{47}
\epsilon&=&\frac{e^2}{a}\left(\frac{1}{2}+\Sigma
+\frac{u^{1/2}}{1+u}\etaup_1\right) \nonumber \\
&=&mc^2\frac{3}{2}\left(\frac{\frac{1}{2}+\Sigma
+\frac{u^{1/2}}{1+u}\etaup_1}{\etaup_1+
\frac{1+u}{2u^{1/2}}\Sigma}\right)\etaup_1
\end{eqnarray}

From (\ref{46}) we see that fixing the angular momentum and charge
determines the relativistic factor. Since $\etaup_1$ is never large
and is only 0.375 in the extreme relativistic limit where $\Sigma$
becomes large the fact that the left hand side is $\simeq
\frac{137}{2}$ shows us that we need the highly relativistic regime
with $u$ close to 1 and $\Sigma\sim 68$. In that regime the terms in
$\Sigma$ dominate both numerator and denominator of the right hand
side of (\ref{47}) so $\epsilon/mc^2\simeq
\frac{3}{2}\etaup_1=\frac{9}{16}$ where we use the limiting value of
$\etaup_1$ for this extremely relativistic case
$\frac{1}{1-u}\sim265,~\gamma\sim530$. Returning to equation(\ref{45})
we find $a=\frac{9}{8} \hbar/mc$ so the radius of the sphere is a
little over $\frac{1}{2\pi}$ Compton wavelengths. The fact that at
this large radius over half the rest mass is in the field energy shows
that it was neglect of the effects of spin and charge polarization
that led to the smaller radius $e^2/mc^2$ being considered
classically.

Is there anything to be learned from these naiive calculations?
Perhaps yes: Firstly polarization charges in the highly relativistic
regime are much higher than the net charge. This helps classical
motions to give angular momenta much higher than $q^2/c$. Carter
already pointed out that $g=2$ for the Kerr metrics in relativity but
did not draw attention to the polarization (Lynden-Bell 2003). $g=2$
is a generic property of all the conformastationary metrics in
General Relativity as well as the Kerr Metrics and Pfister \& King
(2002) suggest it is a rather general result. When the charges are
uniform and fixed to the sphere the $g$ factor varies from 3/2 to 11/6
as $v$ increases (see Appendix I). Secondly the value of the
electromagnetic angular momentum depends on $q^2/c$ and on how
relativistically the system rotates but at given $v/c$; it does not
depend on the radius of the configuration. The electromagnetic field
energy and the dipole moment do depend on size, the latter gives a
size related to the Compton wavelength but although this is two orders
of magnitude larger than the classical radius of the electron the
polarization due to rotation is so strong that more than half the rest
mass can be electromagnetic even at this large radius.

Finally the fact that $\epsilon/mc^2$ is quite near $\frac{1}{2}$ suggests
that a model with a somewhat different structure might yield exactly
$\frac{1}{2}$. If the energy consists of an electromagnetic part
$\epsilon\propto 1/a$ and a string with $\epsilon^\prime =Ta$ then
equilibrium will occur when a $d/da (\epsilon+\epsilon^\prime)=0$
i.e. $\epsilon = \epsilon^\prime$ in which case $mc^2$ would be
$\epsilon +\epsilon^\prime$ and $\epsilon/mc^2=\frac{1}{2}$. We note
that a string loop will have no angular momentum about the axis of the
loop so a purely magnetic angular momentum would then be
correct. However my most recent disc models suggest that {\bf all} the
energy is electromagnetic.

\begin{acknowledgments}
I thank H. Ardavan for his enthusiastic interest in all singularities
due to relativistic motion and for discussion of the convergence of
the solutions near $v=c$ both in the main paper and the Appendix. The
work leading to the approximate solutions with the higher $\delta_n$
neglected was done at the Chateaux de Mons conference in 2001 in
honour of D.O. Gough cf (Lynden-Bell 2003) when it became too hot to
attend the sessions. I thank N.W. Evans for reading and discussing the
manuscript.
\end{acknowledgments}

\appendix
\section{The Rapidly Rotating Uniformly Charged Sphere}

The surface charge $~\sigma=\frac{q}{4\pi a^2}~$ is\; fixed in the
surface. The resulting surface current is $c^{-1}\Omega a
\sin\theta\sigma =J$. We show that a field inside of
$2\boldsymbol{\mu}a^{-3}$ and an external field of a dipole
$(3\boldsymbol{\mu} . \hat{\mathbf r} \hat{\mathbf r}
-\boldsymbol{\mu})/r^3$ satisfies the boundary conditions on the
sphere.

Firstly $B_r =2\mu a^{-3}\cos\theta$ and is continuous across the
sphere from inside to outside.

Secondly $[B_\theta]^{\rm out}_{\rm in}=\mu a^{-3}\sin\theta
(1+2)$ this has to equal $4\pi J=c^{-1}\Omega a^{-1}q\sin\theta$ so
the boundary condition is satisfied provided $\mu =\frac{1}{3}q\Omega
a^2/c=\frac{1}{3}qa\frac{v}{c}$. The electric field is ${\bf
E}=q\hat{\bf r}/r^2$ outside and zero inside.

The angular momentum in the field is found from Poynting's vector

\begin{eqnarray*}
L&=&\frac{1}{4\pi c}\int {\mathbf r}\times({\mathbf E}\times{\mathbf
B})dV = \frac{1}{4\pi c}\int {\mathbf r} \times \left({\mathbf E}
\times\frac{-{\boldsymbol{\mu}}}{r^3}\right)dV \\
&=&\frac{1}{4\pi c} \int \left(\frac{2\boldsymbol{\mu}.\hat{\mathbf{r}}}{r^2}
{\mathbf E} - \frac{q}{r}{\bf B}\right)dV \\
&=&{\bf L}=\frac{\hat{\mathbf z}}{2c} \mu q \int\int (2\cos^2\theta -
3\cos^2\theta +1)\frac{1}{r^4}r^2\sin\theta d\theta dr
\\
&=&\frac{2}{3}\boldsymbol{\mu}q/ac =
\frac{2}{9}~\frac{q^2}{c}~\frac{v}{c} \nonumber 
\end{eqnarray*}
The energy in the electric field is $\frac{1}{2}q^2/a=\epsilon_e$.
The energy in the magnetic field is $\epsilon_m$ where 

\begin{eqnarray*}
\epsilon_m=\frac{1}{8\pi}\int B^2dV
&=&\frac{1}{8\pi}\left[\left(\frac{2\mu}{a^3}\right)^2 \frac{4}{3}\pi
a^3 \right.\\
&+&\left.\mu^2 \int \left(3\cos^2\theta +1\right)r^{-6}dV\right] 
\\
&=&\frac{\mu^2}{a^3}\left(\frac{2}{3}+\frac{1}{3}\right)
=\frac{\mu^2}{a^3} = \frac{1}{9}~\frac{q^2}{a}~\frac{v^2}{c^2}~. 
\end{eqnarray*}
Thus $\epsilon=\epsilon_e+\epsilon_m=\frac{1}{2}~
\frac{q^2}{a}\left(1+\frac{2}{9} ~\frac{v^2}{c^2}\right)~$. The gyromagnetic
ratio is $\mu/L=\frac{3}{2}ac/q$. If all the angular momentum is in
the field then the $g$-factor is

\begin{equation*}
g=\frac{2mc}{q}~\frac{\mu}{L} = 3\frac{mc^2a}{q^2} =
\frac{3}{2}\left(\frac{mc^2}{\epsilon}\right)
\left(1+\frac{2}{9}~\frac{v^2}{c^2}\right)
\end{equation*}

Notice two results of this calculation

\begin{enumerate}
\item
$\frac{2Lc}{q^2}\leqslant \frac{4}{9}$ for $v\leqslant c$ so putting
$L=\frac{\hbar}{c}$ and $q=e$ there could not be an angular momentum
in the field as great as $\frac{\hbar}{2}$. The maximum seems to be
short by the large factor $\sim\frac{9}{4}\times 137$. But the {\bf
conducting} sphere behaves differently.

\item
The $g$ factor is always less than 2 if the energy is entirely
electromagnetic but it is always greater than $3/2$ and when $v=c$ it
becomes $11/6$. Of course if the total energy is not all
electromagnetic this result will change. If we fix the relativistic
factor then $\epsilon \propto 1/a$ if the other energy
$\epsilon_n\propto a^n$ then the condition of equilibrium gives
$\epsilon=n\epsilon_n$ so
$mc^2=\epsilon+\epsilon_n=(1+1/n)\epsilon$. Then the factor
$\frac{mc^2}{\epsilon}=(1+1/n)$.  Cases could be made for $n=1$ a
string holding the electron together, $n=2$ a tension over an area or
$n=3$ a tension throughout a volume so the factor might be 2, 3/2, 4/3
or of course 1 and it is of some interest to decide between these
factors. However in the light of (1) above the insulating sphere with
charge uniformly fixed to its surface is a bad model.
\end{enumerate}

\section{Solution by Generating Functions}

Here we give a second method of solving the mathematical problem which
allows a greater accuracy to be obtained by providing a first order
perturbation treatment to the terms totally neglected in the first
method. Also this second method provides the basis for solving the
problem of the relativistically rotating disk which we treat
elsewhere. We have refrained from replacing the first treatment by the
second because the second, though more powerful, is less direct and
guidance from the first method proved vital in understanding some of
the steps essential to the development of the second.

The equations to be solved are (\ref{5}) and (\ref{6}) and we rewrite
the latter in the form

\begin{equation*}
\sum^\infty_{n=m}(1+\delta_n)\etaup_{2n+1}=\frac{\omega}{4}\Phi_{2m}~.
\end{equation*}
We multiply by $U^m$ and sum from $m=0$ to $\infty$. We define the
generating functions $\Phi(U)$ and $\etaup(U)$ by 

\begin{equation*}
\Phi(U)=\sum^\infty_0\Phi_{2m}U^m ~~ {\rm and} ~~
\etaup(U)=\sum^\infty_{m=0}\etaup_{2m+1}U^m~,
\end{equation*}
and so discover

\begin{equation*}
\sum^\infty_{m=0}\sum^\infty_{n=m}
U^m(1+\delta_n)\etaup_{2n+1}=\frac{\omega}{4}\Phi(U).
\end{equation*}
Now
\begin{equation*}
\sum^\infty_{m=0}\sum^\infty_{n=m}=\sum^\infty_{n=0}\sum^n_{m=0}
\end{equation*}
and
\begin{equation*}
\sum^n_{m=0}U^m=(1-U^{n+1})/(1-U)~,
\end{equation*}
thus
\begin{eqnarray*}
\left[\sum^\infty_{n=0}(1-U^{n+1})\etaup_{2n+1}+\sum^\infty_{n=0}
(1-U^{n+1})\delta_n\etaup_{2n+1}\right] \\
= (1-U)\frac{\omega}{4}\Phi(U)
\end{eqnarray*}
so
\begin{eqnarray}\label{A1}
\eta(1)-U\eta(U)+\sum^\infty_0(1-U^{n+1})\delta_n\etaup_{2n+1}
\nonumber\\
=\frac{\omega}{4}(1-U)\Phi(U)~.
\end{eqnarray}

We now perform similar operations on equation (\ref{5}) multiplying by
$U^m$ and summing from $m=0$ to $\infty$. Reversing the order of
summation
\begin{equation*}
\sum^\infty_{n-1}\sum^{n-1}_{m=0}U^m \Phi_{2n}=\omega\eta (U)~.
\end{equation*}
But
\begin{equation*}
\sum^{n-1}_{m=0}U^m=(1-U^n)/(1-U)~,
\end{equation*}
so
\begin{equation}
\Phi (1)-\Phi(U)=-\omega(1-U)\eta(U)~.
\label{A2}
\end{equation}

Notice that the lack of an $n=0$ term in the sum does not matter as it
cancels out between $\Phi(1)$ and $\Phi(U)$. Putting $U=0$ in the
result gives

\begin{equation}\label{A3}
\Phi(1)-1=-\omega~\eta_1~,
\end{equation}
so we get an expression for $\Phi(U)$ in terms of $\eta$

\begin{equation*}
\Phi(U)=\omega(1-U)\eta(U)+1-\omega~\eta_1~.
\end{equation*}
Inserting this into (\ref{A1}) and solving for $\eta(U)$

\begin{eqnarray}\label{A4}
\eta(U)=\frac{4}{\omega^2Q}\left[\eta(1)+(1-U)\left(\frac{\omega^2}{4}\etaup_1
-\frac{\omega}{4}\right) \right. \nonumber \\
+\left. \sum^\infty_0(1-U^{n+1})\delta_n\etaup_{2n+1}\right]
\end{eqnarray}
where $Q$ is the quadratic obtained from (\ref{12}) by writing $t=-U$

\begin{equation*}
Q=U^2+2(2\omega^{-2}-1)U+1=(U+u)(U+u^{-1})~.
\end{equation*}
In our former approximation we neglected the $\delta_n$ for
$n\geqslant2$ and kept only $\delta_0$ and $\delta_1$. Doing that in
(\ref{A4}) leads, via the method we shall shortly explain, back to our
former solution which gave $\etaup_{2n+1}=C(-u)^n$ for $n\geqslant
2$. We shall adopt this approximate form of solution only for the
small terms with $\delta_n,~n\geqslant 2$ in the sum in (\ref{A4}) but
we shall not assume that the value of $C$ is {\bf exactly} that given
in equations (\ref{25}) and (\ref{26}) and we shall derive refined
values for the $\etaup_{2n+1}$. Thus the sum in (\ref{A4}) becomes
\begin{eqnarray}
\sum^\infty_0(1-U^{n+1})\delta_n\etaup_{2n+1}
=(1-U)\delta_0\etaup_1 + \nonumber\\
+(1-U^2)\delta_1\etaup_3
+C\sum^\infty_2(1-U^{n+1})\delta_n (-u)^n
\label{A5}
\end{eqnarray}
We now define a function $W(t)$

\begin{equation*}
W(t)=\sum^\infty_0\delta_n(-t)^n=\frac{1}{4}\sum^\infty_0
\frac{1}{(2n+1)(2n+2)}(-t)^n~.
\end{equation*}
Writing $w^2$ for $t$ we find

\begin{equation*}
\frac{d}{dw}[wW(w^2)]=\frac{1}{4}\sum\frac{1}{2n+2}(-w^2)^n
\end{equation*}
and
\begin{equation*}
\frac{d}{dw}\left[w^2\frac{d}{dw}(wW)\right] =\frac{w}{4}\sum^\infty_0
(-w^2)^n =\frac{w}{4(1+w^2)}~.
\end{equation*}
Integrating and ensuring convergence at $w=0$ gives us

\begin{equation*}
d/dw(wW)=\frac{1}{8w^2}ln(1+w^2)
\end{equation*}
and integration by parts then gives, again keeping $W$ finite at $w=0$

\begin{equation*}
W=-\frac{1}{8w^2}ln(1+w^2)+\frac{1}{4w}\int^w_0(1+w^2)^{-1}dw~.
\end{equation*}
Replacing $w^2$ by $t$ and taking account of the possibility that $t$
may be negative we find the alternative forms of expression of a
single analytic function 
\begin{equation}\label{A6}
W(t)=-\frac{1}{8t}ln(1+t)+\left\{
\begin{array}{llr}
\frac{1}{4\sqrt{t}}\tan^{-1}\sqrt{t}&t\geqslant 0\\
\frac{1}{8\sqrt{t}}ln\left(\frac{1+\sqrt{-t}}{1-\sqrt{-t}}\right)&t\leqslant 0
\end{array}
\right\}~.
\end{equation}
Notice that the series expansion of the final functions only involve
integer powers of $t$ so the square roots are only involved in the
formal expression in terms of $\tan^{-1} \& \ln$. $W(t)$ is analytic in
the complex $t$ plane in $|t|<1$. Our former approximation retains
only $\delta_0$ and $\delta_1$ so replaces $W$ by the linear
approximation $W_L=\frac{1}{8}-\frac{1}{48}t$.

\noindent At $t=0,~W=W_L=\frac{1}{8},~\Delta W=0$

\noindent At
$t=1,~W=\frac{\pi}{16}-\frac{1}{8}\ln2=0.1097,~W_L=0.1042,~ \Delta
W=0.0055$

\noindent At $t=-1,~W\rightarrow\frac{1}{4}\ln2=0.1733,~W_L=0.1458,\Delta
W=0.0275$

\noindent So while that approximation for W remains good to 5\% in the
more physical range of positive $t$, the accuracy falls to only 13\%
at $t=-1$ and our arguments will involve $W$ there. It is therefore of
importance to get the more accurate result to which this Appendix is
devoted. Re-expressing (\ref{A5}) in terms of $W$

\begin{eqnarray*}
\sum^\infty_0(1-U^{n+1})\delta_n\etaup_{2n+1}=(1-U)\delta_0(\etaup_1-C)+\\
+ (1-U^2)\delta_1(\etaup_3 +Cu)+C[W(u)-UW(uU)]
\end{eqnarray*}
so our expression (\ref{A4}) for $\eta(U)$ becomes
\newline

$\eta(U)=$
\begin{eqnarray}\label{A7}
\frac{4}{\omega^2Q} \{ \eta(1)+(1-U)\left[
\frac{\omega^2}{4}\etaup_1
-\frac{\omega}{4}+\delta_0(\etaup_1-C)\right]+ \nonumber \\
\left. (1-U^2)\delta_1(\etaup_3+Cu)+  C [W(u)-UW(uU)]\right\}~.
\end{eqnarray}

$\etaup_{2l+1}$ is the coefficient of $U^l$ in this re-expressed as a
power series in $U$. However in our solution (\ref{11}) of the
recurrence relation (\ref{10}) we had to choose the convergent
solution and throw away the divergent one and only then did we have
enough equations to yield a solution. We now have to find the
analogous step in our new approach. The numerator of $\eta(U)$ is
analytic everywhere that $W(uU)$ is analytic i.e. within
$|U|=u^{-1}$. Thus any poles that can give us divergences in the high
order coefficients of $U^l$ can only come from the zeros of $Q$. One
such pole is at $U=-u^{-1}$ which is outside the $|U|=1$ circle,
inside which $\eta(U)$ should be analytic, but the other is within
that circle at $U=u$. It is this pole that leads to the divergent
solution of the recurrence relation whose coefficient $D$ had to be
chosen as zero in (\ref{11}). Thus the requirement of convergence in
our coefficients $\etaup_{2n+1}$ can be replaced by the requirement
that $\eta(U)$ should be analytic with no poles within the circle
$|U|=1$. This in turn translates into the requirement that the
numerator of $\eta(U)$ must vanish at $U=-u$ so
\begin{eqnarray*}
\eta(1)+(1+u)\left[\frac{\omega^2}{4}\etaup_1 -\frac{\omega}{4}+\delta_0
(\etaup_1 -C)\right] + \\
+(1-u^2)\delta_1(\etaup_3+Cu)+C[W(u)+uW(-u^2)]=0
\end{eqnarray*}
without first having the solution by the recurrence relationship it
would have been difficult to see this requirement. We now use this
form of $\eta(1)$ in (\ref{A7}). At the same time we replace $\omega^2,Q$
by their values in terms of $u$ given in equation (\ref{13}). Then 

\begin{eqnarray}\label{A8}
\eta(U)&=&(1+Uu)^{-1} 
\left[
-u\eta_1+\mbox{$\frac{1}{2}$}u^{\frac{1}{2}}(1+u)+ \right. \nonumber \\
&+&\left. (1+u)^2
\left\{-\delta_0(\etaup_1-C)+(u-U)\delta_1(\etaup_3+Cu)- \right. \right. \nonumber \\
&-&\left. \left. C[uW(-u^2)+UW(uU)]/(u+U) \right\} \right]
\end{eqnarray}
Now $\eta(o)=\etaup_1$ hence
\begin{eqnarray}\label{A9}
\etaup_1&=&- u\eta_1 +\frac{1}{2}u^{1/2}(1+u) +(1+u)^2
\{-\delta_0(\etaup_1-C)+ \nonumber \\
&+ & u\delta_1(\etaup_3+Cu)-CW(-u^2)\}
\end{eqnarray}

We use this expression to simplify (\ref{A8}) into
\begin{eqnarray}\label{A10}
\eta(U)&=&(1+Uu)^{-1}\{\etaup_1+(1+u)^2U \times  \nonumber \\
&\times&\left.\hspace{-2mm} \left[-\delta_1(\etaup_3+Cu)\hspace{-1mm}
+\hspace{-1mm}C\frac{W(-u^2)\hspace{-1mm}-\hspace{-1mm}W(uU)}{U+u}\right]\right\}~.
\end{eqnarray}

Notice that the term involving the $Ws$ has no singularity when $U=-u$
and that term in the overall numerator takes the value
\begin{equation*}
-u[W(-u^2)-W(-1)]/(1-u^2)
\end{equation*}
when $U=-1/u$. This is finite for $u<1$. Thus isolating the term with
a pole at $U=-1/u$ we have
\begin{eqnarray}\label{A11}
\eta(U)&=&\frac{C}{1+Uu} - 
\frac{(1+u)^2}{u}\delta_1 (\etaup_3 +Cu) + C(1+u)^2\times \nonumber \\
&\times& \hspace{-2mm}\left\{\frac{U\left[\frac{W(-u^2)-W(uU)}{U+u}\right]\hspace{-1mm}-\hspace{-1mm}
\left[\frac{W(-u^2)-W(-1)}{1-u^2}\right]}{1+Uu}\right\}
\end{eqnarray}
where the numerator of the last term is zero at $U=-1/u$ so it has no
pole there either. The first pole comes from the first term which
gives the asymptotic expansion $C\Sigma(-Uu)^n$ as expected from the
asymptotic form of the recurrence relation. This justifies the
nomenclature of calling its coefficient $C$ which was shorthand for
(cf (\ref{A10}))
\begin{eqnarray}\label{A12}
C=\etaup_1+(1+u)^2u^{-1}\times \nonumber \\
\left\{\delta_1(\etaup_3+Cu) +CU
\frac{W(-u^2)-W(-1)}{1-u^2}\right\}~.
\end{eqnarray}
Now from the definition of $\eta(U)$ we have
$\eta^\prime(0)=\etaup_3$. Applying this to (\ref{A11}) we find the
third equation which together with (\ref{A9}) and (\ref{A12}) we solve
for $\etaup_1,~\etaup_3$ and C
\begin{eqnarray}\label{A13}
\etaup_3=-Cu +C(1+u)^2\times \nonumber \\
\left[u\frac{W(-u^2)-W(-1)}{1-u^2} +
\frac{W(-u^2)-W(0)}{u}\right] = \nonumber \\
-Cu+Cu (1+u)^2g_0
\end{eqnarray}
The resulting value of $\etaup_3+Cu$ we substitute into (\ref{A12}) to
obtain

\begin{equation}\label{A14}
\etaup_1 =C\{1-(1+u)^2[g_1 +\delta_1 g_0(1+u)^2]\}
\end{equation}
where $g_1=\frac{W(-u^2)-W(-1)}{1-u^2}$ and
$g_0=g_1+\frac{W(-u^2)-W(0)}{u^2}$ when only $\delta_0$ and $\delta_1$
are retained $g_1=-\delta_1$ and $g_0$ is zero.

We now use (\ref{A13}) and (\ref{A14}) in (\ref{A9}) to obtain an
expression for $C$

\begin{equation*}
C=\frac{1}{2}u^{1/2}/\Delta
\end{equation*}
where

\begin{eqnarray}\label{A15}
\Delta &=& 1-(1+u)^2\{g_1[1+(1+u)\delta_0] +\nonumber \\
&+&(1+u)\delta_1g_0 [1+u+u^2)
+(1+u)^2\delta_0]\} + \nonumber \\
&+&(1+u)W(-u^2)
\end{eqnarray}
which reduces to (\ref{26}) in the appropriate limit. With
$\etaup_1~\etaup_3$ and $C$ now determined $\eta(U)$ is known from
(\ref{A11}).

We are now in a position to obtain a more accurate value for the
surface density of charge at the poles and the equatorial velocity at
which it changes sign. According to (\ref{1}) we need 

\begin{equation*}
\sum^\infty_0(4n+1)\Phi_{2n}=4\Phi^\prime(1)+\Phi(1)~.
\end{equation*}
But by (\ref{A2}) $\Phi^\prime(1)=-\omega\eta(1)$

\noindent and by (\ref{A3}) $\Phi(1)=1-\omega\eta_1$

\noindent so $\sigma_p =\frac{q}{4\pi a^2}\{1-\omega[\etaup_1+4\eta(1)]\}$

Evaluating this expression as a function of $u$ we find that
$\sigma_p=0$ when $u=0.455$ which is in better than expected agreement
with the crude calculation of section 4 which gave $u=.454$ or 93\% of
$c$ for the equatorial speed.

We now have our solution: $C$ is given by (\ref{A14}) and
(\ref{A15}). Let us recap the notation $\omega=\Omega a/c=\frac{v}{c}$

\begin{eqnarray*}
u&=& 2\omega^{-2}\left[1-\frac{1}{2}\omega^2-\sqrt{1-\omega^2}\right]~,\\
{\rm so}~~~\gamma&=& \left(1-\frac{v^2}{c^2}\right)^{-1/2}=\frac{1+u}{1-u}
\end{eqnarray*}
$W(t)$ is the function defined by (\ref{A5})

\noindent $g_1$ is the expression $[W(-u^2)-W(-1)]/(1-u^2)$

\noindent $g_0$ is $g_1+u^{-2}[W(-u^2)-W(0)]$

\noindent $\gamma_0=W(0)=\frac{1}{8}~,\qquad \delta_1=\frac{1}{48}$

\noindent Thus all terms in $C$ are known once $\omega$ or $u$ are
given. (\ref{A13}) then gives $\etaup_1$ and (\ref{A12}) gives
$\etaup_3$.

\noindent The function $\eta(U)$ is now given by (\ref{A7}) with all
terms therein known.

The coefficient of $U^n$ in the expansion of $\eta(U)$ as a power
series in $U$ is our refined solution for $\etaup_{2n+1}$ but with the
first two $\etaup_1$ and $\etaup_3$ now explicitly known the others
may also be found by direct repeated application of the exact
recurrence relation (\ref{10}) e.g.

\begin{equation*}
\etaup_5 =C\left\{u^2+(1+u)^2[g_1+\delta_1-g_0(1+u)^2]\right\}
\end{equation*}

In practice the most interesting region is the relativistic
region. $u=\frac{1}{2}$ gives $\gamma=3$, $u=0.8$ gives $\gamma=9$
while $u=0.9$ gives $\gamma=19$. Evaluating $W(-u^2)$ for such values
we find

\begin{eqnarray*}
W(-u^2)=\frac{1}{8u^2}\left[(1-u)\ln(1-u)+(1+u)\ln(1+u)\right] \\
W(-1)=\mbox{$\frac{1}{4}$}\ln 2=.173,~~W(0)=\mbox{$\frac{1}{8}$}
\end{eqnarray*}
Near $u=1~g_1=+\frac{1}{8u^2(1+u)}\times$
\begin{eqnarray*}
\times[\ln(1-u)-\left(\ln 2 - 1\right) &+& (1-u)(\ln
2-\mbox{$\frac{1}{4}$})] \\
g_0 &-& g_1 =\\
\mbox{$\frac{1}{8}$}u^{-4}[(1-u)\ln(1-u) &+& (1+u)\ln(1+u)-u^2]
\end{eqnarray*}
the dominant term is the $\ln(1-u)$ term in $g_1$.

Notice that near $u=1$ these terms become large which invalidates our
perturbation theory beyond $\gamma\sim 20$. The $\gamma >20$ region is
treated differently in the main body of the paper.


\begin{thebibliography}{}

\bibitem{1}
Carter B 1968a Phys Rev 174, 1559

\bibitem{2}
Carter B 1968b Commun Math Phys 10, 280

\bibitem{3}
Chandrasekhar S 1976 Proc R Soc London A 349, 571

\bibitem{4}
Lynden-Bell D (2000) MNRAS 312, 301

\bibitem{5}
Lynden-Bell D (2003) Stellar Astrophysical Fluid Dynamics pp 369-375
Eds M.J. Thompson \& J Christiansen-Dalsgaard CUP = astro-ph/0207064

\bibitem{6}
Newman E T (1973) J Math Phys 14, 102

\bibitem{7}
Newman E T (2002) Phys Rev D 65, 104005

\bibitem{8}
Newman E T Couch E Channapared K Exton A Prakesh A Torrance R (1965) J
Math Phys {\bf 6}, 918

\bibitem{9}
Page D N 1976 Phys Rev D 14, 1509

\bibitem{10}
Pfister H \& King M 2002 PACS 09.40 Nr, 13-40 EM

\end{thebibliography}
\end{document}